# Magnetism of a rhombohedral-type pyrochlore-derived Kagome series, $Mn_2R_3Sb_3O_{14}$ (R= Rare-earths)

Venkatesh Chandragiri, Kartik K Iyer, K. Maiti, and E.V. Sampathkumaran
*Department of Condensed Matter Physics and Materials' Science, Tata Institute of Fundamental Research, Homi Bhabha Road, Colaba, Mumbai 400005, India*

**Abstract**

The results of magnetic investigations on a new series of compounds, $Mn_2R_3Sb_3O_{14}$, containing 2-dimensional Kagome lattice of R ions and belonging to pyrochlore family, are presented. Crystallographic features of light R members (R= La, Pr, and Nd) of this family, as established in the recent literature, have been reported to be novel in many aspects, in particular, the rhombohedral nature of the structure which is rare among pyrochlores. It was also reported that, as the R becomes heavier, beyond R= Sm, the fraction of well-known cubic pyrochlore phase tends to gradually dominate. Here, we report that we are able to form the Gd member in the rhombohedral form without noticeable admixture from the cubic phase. With respect to magnetic behavior, our magnetization measurements on the La member reveal that Mn exists in divalent state without any evidence for long range magnetic ordering down to 2 K, a behavior (that is, suppressed magnetism) which is not so common for Mn based oxides, though antiferromagnetism below 2 K is not ruled out. Nd and Gd members are, however, found to show distinct features above 2 K in magnetic susceptibility and heat-capacity, attributable to long-range magnetic ordering from respective rare-earth sublattice. The experimental results with respect to magnetism are found to be consistent with the results from *ab initio* band structure calculations performed for the La case. The calculations imply that electron correlation is important to describe insulating behavior.

# 1. INTRODUCTION

The magnetic behavior of manganese containing compounds in metallic and nonmetallic environment have been attracting considerable attention in recent years. Thus, for instance, apart from well-known colossal magnetoresistance perovskite-based manganites [1], Mn pnictides LaMnPO [2], $CaMn_2As_2$ [3], and $BaMn_2As_2$ [4] which are insulators due to strong Coulomb interaction, and metals like binary Mn compounds [5], $RMn_2X_2$ (R= Rare-earths; X= Si, Ge) [6] exhibiting pronounced magnetic character due to Mn, reduced moment metallic systems like MnSi [7], $YMn_2$ [8], $HfMnGa_2$ [9], and most recently $CaMn_2Al_{10}$ on the verge of magnetic order [10], are of great interest. It is worthwhile to note that Mn tends to order magnetically at rather high temperatures, even in a family when other transition metals like Fe, Co, Ni and Cu do not exhibit magnetic moment, as demonstrated for $RMn_2X_2$ family [6]. Clearly, it is worthwhile to search for new manganese-based systems and to study their magnetic behavior.

The oxide compounds with the stoichiometry 2:2:7 (of the type $A_2B_2O_7$) attracted considerable attention in condensed matter physics [11 - 14]. In general, three kinds of structures are commonly known for this composition, namely, cubic pyrochlores, orthorhombic weberite, and the trigonal weberite, and the corresponding space groups are $Fd\bar{3}m$, $Imma$ and $P3_121$ respectively. The relationship between these structures has been discussed by Kai and Nino [15], and Fu and Ijdo [16]. It was recently reported [16] that another family of pyrochlores of the type, $Mn_2R_3Sb_3O_{14}$, however, in a rhombohedral structure (see figure 1) in the space group, $R\bar{3}m$, not commonly known for pyrochlores, can be synthesized. In figure 1, we show the crystallographic features in various orientations, with metallic ions only, omitting oxygen atoms for the sake of clarity. It is clear from figure 1 that the structure is a 2-dimensional rare-earth Kagome lattice. This new structure differs from the common cubic pyrochlores in the following sense: The former consists of fully ordered Mn:R in the A-sites and Mn:Sb in the B sites with the ratio 1:3, running along the four equivalent [111] directions of cubic pyrochlore; one of the four [111] planes of the pyrochlore naturally becomes trigonal $c$-axis of the rhombohedral structure. In the present family of compounds, $R_3Mn$ slabs contain distorted edge-sharing $RO_8$ cubes forming hexagonal rings with $Mn1O_6$ octahedra at the center; likewise, $Sb_3Mn$ slabs contain hexagonal rings of Sb (octahedra) with $Mn2O_8$ (with coordination with 8 oxygens) at the center. The lifting of cubic symmetry is attributed to cation ordering, and the rhomobohedral form is left with only one 3-fold axis along the [111] direction of cubic pyrochlore form. There are three non-equivalent oxygen sites, which are denoted as O1, O2, and O3 in this paper. It was also reported that the family forms in single phase for the beginning of the rare-earth series only (R= La, Pr and Nd), while the attempts to prepare heavier members (beyond Sm) resulted in a multiphase with increasing dominance of cubic pyrochlore phase.

The aim of the present investigation is to (i) attempt to synthesize some other members of this family, viz., R= Eu, Gd, and (ii) to understand magnetism of this novel family of insulating oxides by carrying out magnetization (M) and heat-capacity (C) studies as well as *ab initio* band structure calculations.

# 2. EXPERIMENTAL DETAILS

Attempts were made to synthesize polycrystalline samples for R= La, Pr, Nd, Eu, Gd, Tb, Dy, Ho and Er, as described in Ref. 1, starting from respective oxides: $MnO_2$, $R_2O_3$, $Pr_6O_{11}$, $Tb_4O_7$ and $Sb_2O_3$. These oxides in proper proportions were thoroughly mixed, and heated at 1073 K in an alumina crucible for 1 day. Subsequently, after grinding, the specimens were pelletized, placed in Pt foil and heated at 1473 K with intermediate grindings a few times. The specimens were characterized by x-ray diffraction (Cu $K_\alpha$). Magnetization measurements as a function of temperature ($T$= 2-300 K) and magnetic field ($H$) were performed employing a commercial magnetometer. Heat-capacity data were also collected with the help of a Physical Properties Measurements System (Quantum Design).

The energy band structure of the La compound, $Mn_2La_3Sb_3O_{14}$, in this family was calculated using full potential linearized augmented plane wave (FLAPW) method as adapted in Wien2k software, within the local spin density approximations (LSDA). The crystal structure parameters were fixed to the lattice parameters found in the room temperature *x*-ray diffraction results. For La case, $a = b = 7.539$ Å and $c =$ 17.789 Å. The muffin-tin radii, $R_{MT}$ for Mn, La, Sb and O were set to 1.94 a.u., 2.35 a.u., 2.06 a.u. and 1.67 a.u., respectively. The convergence for different calculations were achieved considering 1000 $k$ points within the Brillouin zone. The error bar for the energy convergence was set to < 1 meV.

## 3. RESULTS AND DISCUSSIONS

X-ray diffraction patterns are shown in figure 2. As known earlier [12], the diffraction patterns for La, Pr and Nd could be indexed to the rhombohedral structure, and for heavier R members like Eu, Tb, Dy, Ho and Er, the cubic phase (marked by * in figure 2) starts appearing gradually with increasing atomic number of R with the least intensity for Eu among the compounds attempted in this study. It appears that no attempt was made to synthesize this compound in Ref. 12. What is surprising is that, despite difficulties in forming Eu compound (a lighter R member) in single phase, the heavier rare-earth member, R= Gd, forms in the rhombohedral structure apparently without any indication for cubic phase within the detection limit of $x$-ray diffraction (<2%). In order to avoid difficulties in interpreting the results due to the interference from the presence of the cubic phase, magnetic studies are restricted to single phase rhombohedral compounds only in this investigation, the results of which presented below.

In the mainframe of figure 3, inverse susceptibility ($\chi^{-1}$) obtained in a field of 5 kOe is plotted. In order to obtain magnetic behavior of Mn, the behavior of La compound is useful. The plot of $\chi^{-1}$ versus $T$ is found to be linear above 150 K (as shown by the continuous line in figure 3) and the paramagnetic Curie temperature ($\theta_p$) and effective moment ($\mu_{eff}$) obtained from this linear region are found to be about +10 K and 5.74 $\mu_B$/Mn respectively. The value of $\mu_{eff}$ is marginally lower than that expected (5.92 $\mu_B$/Mn for high-spin (S= 5/2) of divalent Mn. There is a deviation from this high temperature linearity below 150 K. Though the origin of this is not clear to us, the role of gradual dominance of short range magnetic correlations with decreasing temperature is not ruled out. A linear extrapolation of the plot below 150 K intercepts $x$-axis around -10 K, as though antiferromagnetic correlations tend to dominate with decreasing temperature. A support for antiferromagnetism in the ground state is provided by band structure calculations presented in this article. There is no evidence for any feature attributable to long-range magnetic ordering in the $\chi(T)$ data down to 2 K measured, despite reasonably large values of $\theta_p$, for both the coordinations of Mn. It is therefore of interest to focus on future studies to understand whether this compound belongs to the family of 'spin-liquids'.

The plots of $\chi^{-1}$ are found to be linear over a wide $T$-range above 10 K for Pr, Nd, and Gd members (figure 3). The values of $\theta_p$ as obtained from Curie-Weiss fitting in the range 150-300 K are found to be ~ -22 K, -20 K and -4 K respectively. The higher $\theta_p$ values for lighter rare-earths compared to the case of Gd compound implies decreasing role of 4$f$ – conduction electron hybridization on the exchange interaction strength and/or increasing role of ferromagnetic exchange over antiferromagnetic correlations as one moves from Pr to Gd. In the case of such rare-earth containing compounds, it is customary in the literature to assume the free ion values of $\mu_{eff}$ for R ions, due to localized nature of 4$f$ electrons, to derive $\mu_{eff}$ for other elements in the compound. With this procedure, we obtained $\mu_{eff}$ values of Mn in all these compounds, which turn out to be close (within 0.2 $\mu_B$/Mn) to the value obtained for the La analogue. As the temperature is lowered, some new features emerge, which can be attributed to magnetic ordering, for Nd and Gd members only (but not for Pr) in the vicinity of 2.5 and 8 K in the plots of $\chi$ versus $T$ (see the insets of figure 3); though there is some degree of bifurcation of zero-field-cooled (ZFC) and field-cooled (FC) curves, measured in 100 Oe (at least for Nd and Gd cases), the observed irreversibility is completely different from that expected for spin-glass freezing, as ZFC curves do not show a cusp. In fact, we additionally performed isothermal remnant magnetization studies at 2 K as well as ac $\chi$ measurements for the Gd case and these data are not supportive of spin-glass freezing.

We have also measured isothermal magnetization as a function of $T$ up to 80 kOe as well as low-field hysteresis loops at 1.8 K (figure 4). In the case of La sample, the plot of M(H) is linear till rather high fields. However, there is only a weak curvature near very high fields, whereas a paramagnetic state (derived from Brillouin function for 2 K for spin= 5/2) would yield a different plot with a tendency towards saturation at lower fields. This means the existence of short range antiferromagnetic correlations or a precursor to the onset of long range antiferromagnetic order from Mn presumably setting in well below 2 K (as indicated by heat-capacity data discussed below). For $R$ = Pr and Nd, we do not find any evidence for complete saturation till high fields. Linearity of the *M(H)* plot over a wide field range for Pr case is consistent with the absence of long-range magnetic ordering and the resemblance of the curve to that of La compound implies the existence of short-range magnetic correlations among Mn ions. The convex curvature of the plot for Nd case could be due to magnetic ordering, presumably from Nd ions (also see heat capacity behavior, discussed below). Naturally, short range magnetic correlations among Mn, if it persists as in the La case, should be antiferromagnetic. In the case of Gd compound, *M(H)* tends towards saturation above about 40 kOe and the linear extrapolation of high-field values to zero-field yields about 20 $\mu_B$/formula unit, which is about 1 $\mu_B$ below that expected for 3 Gd ions (21 $\mu_B$/formula unit). This finding signals that Gd could order ferromagnetically, but the antiferromagnetic Mn sublattice could be oriented in such a way (e.g., by canting) to lower the net saturation moment per formula unit. Low field hysteresis plots at 1.8 K are hysteretic for Nd and Gd cases with low coercive fields, suggesting thereby the existence of a ferromagnetic component for both. It may be mentioned that, in the case of Gd, the virgin curve lies outside the envelope curve; such a behavior is commonly known for disorder-broadened first-order transitions as well as for some other magnetic systems, e.g. $Fe_2O_3$ [17]. The results overall suggest that the magnetic ordering behavior of these compounds could be quite complex.

In order to support conclusions with respect to the magnetic ordering behavior, we show the plots of *C(T)* in figure 5. For the Nd case, in zero field, there is an upturn below about 7 K followed by a prominent peak at about 2.5 K; corresponding features appear around 10 K and 6 K for Gd case. These results confirm bulk nature of magnetic ordering (that is, not due to possible traces of magnetic impurities), also ruling out spin-glass freezing (in which case the features at the transition would be smeared). These peaks shift towards higher temperature range with increasing *H*. This finding supports possible dominant ferromagnetism of R ions at low temperatures. For the La and Pr cases, we could not observe any peak, consistent with the conclusion on the absence of long range magnetic ordering above 2 K, made above. However, the plots of C/T shown in the inset exhibit a weak upturn below about 4 K. The low-temperature upturn is visible even in the plots of C vs *T* itself for the La case. This finding implies that there could be a peak due to a magnetic transition below 2 K in this case. This upturn in C vs *T* is suppressed for 50 kOe (see the curve for the La case), which signals the magnetic transition, if present, may be antiferromagnetic, and not ferromagnetic. This may also be consistent with negative $\theta_p$ at low temperatures. It is attributable to magnetic ordering from Mn sublattice occuring well below 2 K. It appears that there is a magnetic-field dependence of the values above 2 K (see the curve 50 kOe) in the range 2-15 K, which can be due to short-range magnetic correlations. This finding could be consistent with the fact that the plot C/T versus $T^2$ below 20 K does not extrapolate to origin expected for insulators. Such enhanced linear terms in heat-capacity have been known to arise from magnetic precursor effects, e.g., critical spin fluctuations extending over an unusually wide temperature range [18].

In order to investigate the existence of such magnetic interactions within the Mn sublattice, we have calculated the energy band structure of the La compound, $Mn_2La_3Sb_3O_{14}$ using density funtional theory [19]. We observe that spin-polarized calculation converges to a much lower energy ground state (~ 13 eV) than the non-magnetic case. Interestingly, antiparallel coupling between Mn1 and Mn2 sites corresponds to the lowest energy case with the parallel coupling appearing about 60 meV higher in energy. We also find that the ground state is metallic in all the magnetic configurations studied. However, our electrical resistivity studies revealed that these compounds are highly insulating. It is now established in the literature, based on studies on many other oxide systems, that this is normal for 3d electron systems and that inclusion of electron correlation [20] among the Mn *d*-electrons is essential to capture the exact ground state properties, although the magnetic phase has often been found to be captured well in these calculations [21].

The calculated partial density of states (PDOS) for the antiferromagnetic ground state of $La_3Mn_2Sb_3O_{14}$ is shown in Fig. 6. The up and down spin density is shown with positive and negative scale in the *y*-axis respectively. The antiferromagnetic alignment of Mn1 and Mn2 is evident in the figure 6(a) showing Mn 3*d* PDOS with majority spin density of states having opposite sign. Both Mn1 and Mn2 exhibit large exchange interaction strength (~ 4 eV), which is much larger than the crystal field splitting (of about 2 eV, see figure 6a) found in this system. While such a trend is observed in other Mn oxides, the exchange coupling strength in this system seems to be significantly higher than in other oxides. The major contribution from Mn 3*d* PDOS appears above -2 eV energy. The O 2*p* PDOS shown in Fig. 6(b) primarily contributes below -2.5 eV. Significant contribution from Mn 3*d* PDOS in this energy range and O 2*p* contribution in the Mn 3*d* PDOS range suggest strong Mn 3*d* – O 2*p* hybridization; the states below -2.5 eV possess bonding character and the ones above -2 eV are the antibonding states.

Sb 5*p* and 5*s* contributions are shown in Fig. 6(c). 5p PDOS appears primarily in two energy ranges; -8.5 to -5 eV and 2.5 to 11 eV, with the major contribution above the Fermi level. No contribution is observed at the Fermi level. On the other hand, 5*s* PDOS exhibits significant contribution at the Fermi level along with a large contribution below -8 eV energy. Interestingly, O3 2*p* PDOS exhibits significant contribution in the Sb 5*s* and 5*p* PDOS regimes indicating hybridization between the two. No contribution is observed in the Mn 3*d* PDOS region suggesting insignificant interaction between the Mn and Sb entities in these materials. The electronic states possessing La orbital character primarily appear above the Fermi level along with a small 5*d* contribution in the vicinity of -5 eV energy as a signature of hybridization with the O2 2*p* states.

In Fig. 7, we show the PDOS corresponding to Mn 3*d* spin-orbitals separately for both Mn1 and Mn2. Mn1 forming the MnO$_6$ distorted octahedra exhibits distinct behavior of the electronic states with different symmetry. *d*$_{xy}$, *d*$_{xz}$ and *d*$_{yz}$ orbitals are degenerate and exhibit two sets of narrow energy bands of width of about 1eV for both up and down spin states. *d*$_{x2-y2}$ and *d*$_{z2}$ bands are also narrow and appear almost in the same energy range with somewhat larger gap between the two bands. The up spin states are close to complete occupancy and the down spin states are completely empty. On the other hand, the 3*d* states corresponding to Mn2 exhibit significantly different scenario, although the width of the energy bands are almost about 1 eV as found in the case of Mn1. The contribution at the Fermi level is very very weak and appear only due to 3*d*$_{z2}$ states with all other states exhibiting large gap of about 2.5 eV. These results indicate that although the Mn 3*d* – O 2*p* hybridization is significant, the valence electrons possess strong local character even within the local density approximation treatment. Near complete occupancy of the Mn 3*d* up-spin states supports essentially divalent character of Mn as found experimentaly.

## 4. CONCLUSIONS

We establish that the rarely occurring rhombohedral form [22] of pyrochlores extends to Gd as well in the newly discovered [16] series, Mn$_2$R$_3$Sb$_3$O$_{14}$, with R-based 2-dimensional Kagome layers. Our magnetization and heat capacity studies reveal that Mn does not exhibit long range magnetic order down to 1.8 K in light rare-earths, although some signatures of inter-site antiferromagnetic exchange interaction between Mn (as inferred from La case) is present at low temperatures. This finding for Mn compounds is interesting, as it is not common that the long-range magnetic ordering temperature of large magnetic-moment containing divalent Mn ion is suppressed to this extent. Nd and Gd members order magnetically, possibly with a dominating ferromagnetic component, at low temperatures. It is interesting that these R ions do not undergo spin-glass freezing, despite that the Kagome-lattice arrangement favors geometrically frustrated magnetism in the event of antiferromagnetic interaction; this observation is consistent with the proposal that ferromagnetic intersite interaction tends to dominate over antiferromagnetic interaction. The results from band structure calculations of the La compound representing the magnetism of Mn sublattice are found to be consistent with the experimental results and indicate importance of correlation physics in these materials. Finally, no evidence for magnetodielectric or magnetoelectric coupling is observed in these compounds.

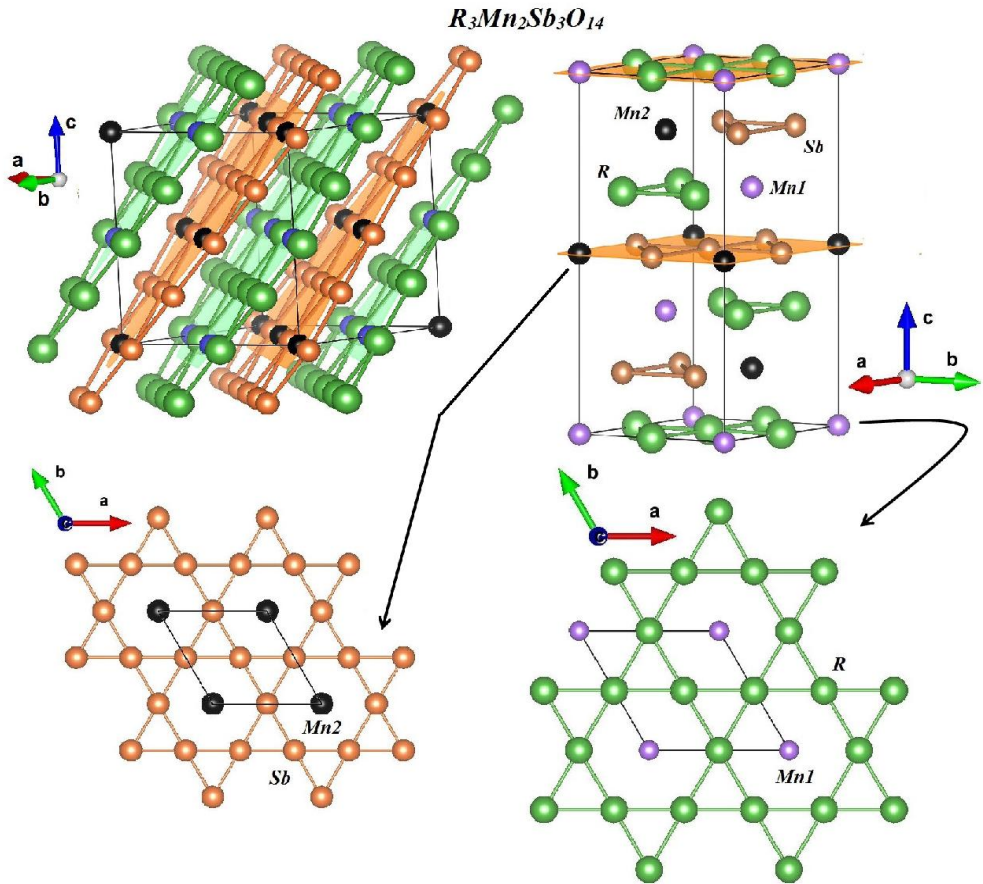

Figure 1: Crystal structure of Mn$_2$R$_3$Sb$_3$O$_{14}$ looking from various directions, omitting oxygen atoms for the sake of clarity. **Top left:** Layers of R$_3$Mn and Sb$_3$Mn shown in the cubic pyrochlore form. **Top right**: Rhombohedral unit cell of the compounds. **Bottom left**: Sb$_3$Mn slab viewed along *c*-axis. **Bottom right:** R$_3$Mn slab viewed along *c*-axis. For the sake of clarity, oxygen atoms in Mn1O$_6$, Mn2O$_8$, RO$_8$, and SbO$_6$ polyhedra are omitted.

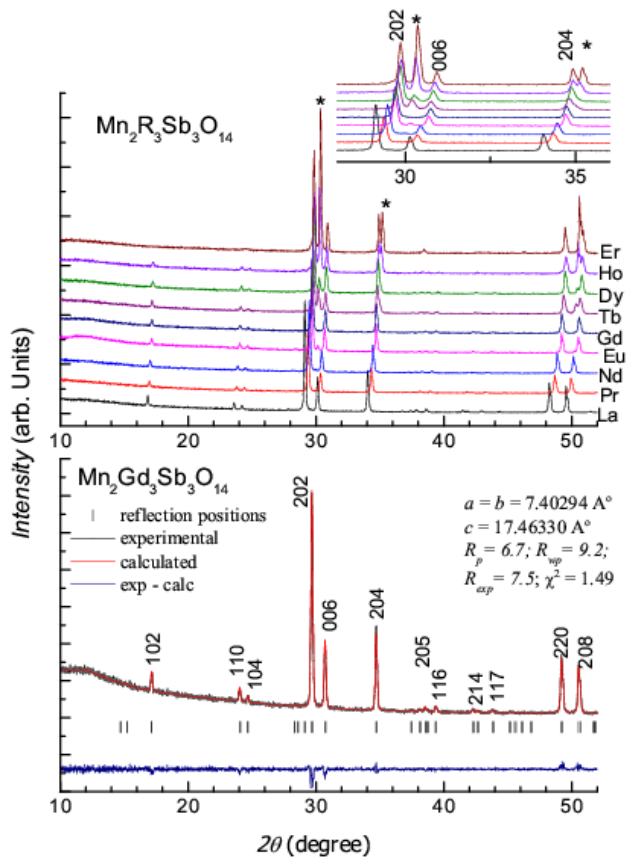

**FIGURE 2:** **(Top)** X-ray diffraction patterns of $Mn_2R_3Sb_3O_{14}$ (R= La, Pr, Nd, Eu, Gd, Tb, Dy, Ho, and Er) compounds. The peaks with asterisk correspond to the cubic phase. Inset shows patterns in expanded form to show changes in lattice constants consistent with lanthanide contraction. **(Bottom)** X-ray diffraction pattern for Gd sample with Reitveld fitting.

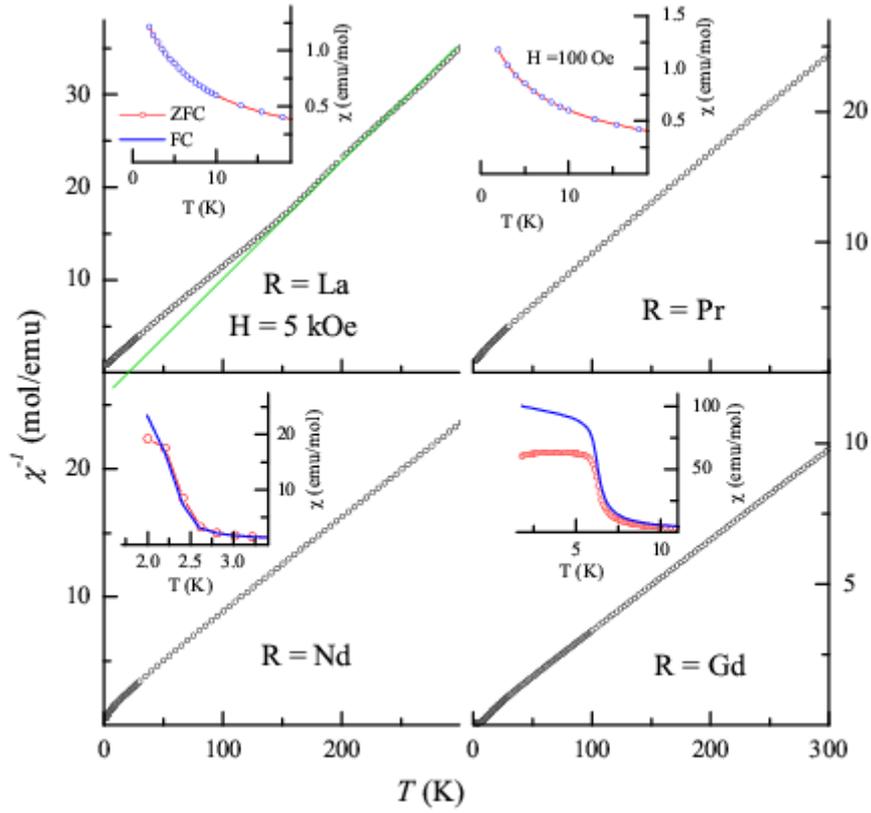

**FIGURE 3:** Inverse susceptibility as a function of temperature for $Mn_2R_3Sb_3O_{14}$ compounds, measured in a field of 5 kOe. Insets show $\chi$ behavior for ZFC (points) and FC (continuous red lines) conditions, measured in a field of 100 Oe, to highlight magnetic ordering behavior. A line (in colour) is drawn though the Curie-Weiss region at higher temperatures for the La case.

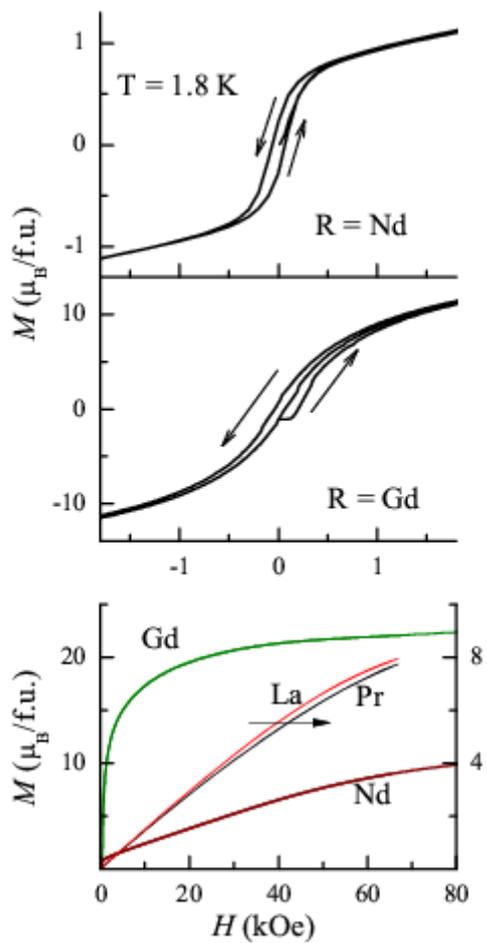

**FIGURE 4:** Low-field magnetic hysteresis loops (upper figure) and isothermal magnetization behavior (bottom figure) at 1.8 K for $Mn_2R_3Sb_3O_{14}$. The density of data points is so high that the lines through the data points only are shown.

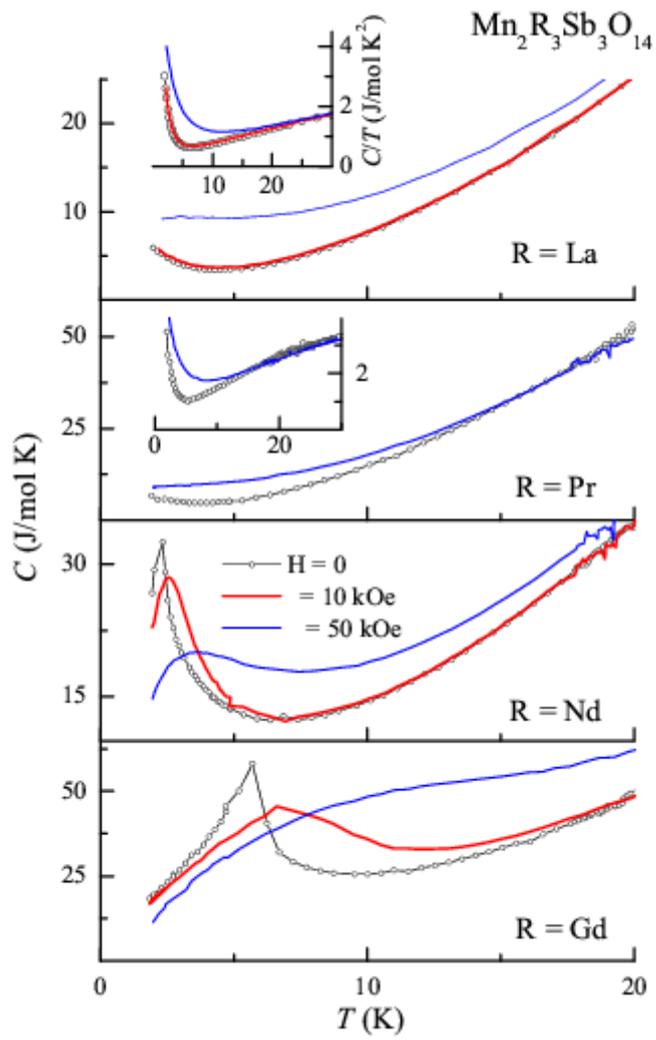

**FIGURE 5:** Heat-capacity as a function of temperature for $Mn_2R_3Sb_3O_{14}$ compounds in the absence of external field as well as in 10 and 50 kOe. The data points are shown for zero-field only, while for other fields, only lines drawn through the data points are shown omitting data points, for the sake of clarity.

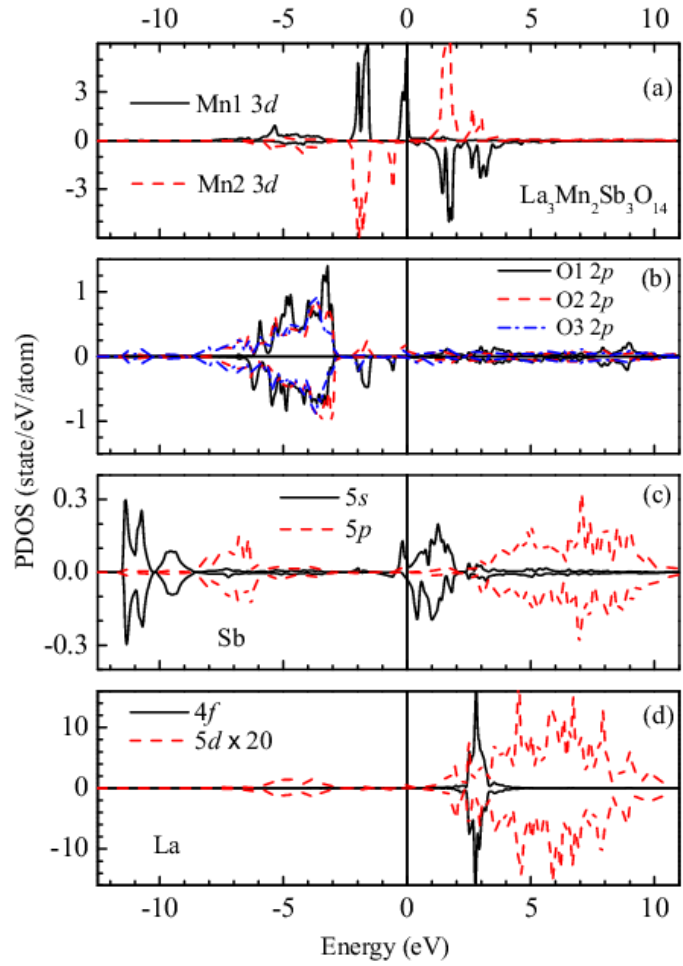

Fig 6: Calculated partial density of states (PDOS) of (a) Mn1 and Mn2 3$d$, (b) O1, O2 and O3 2$p$, (c) Sb 5$s$ and 5$p$ and (d) La 4$f$ and 5$d$ in La$_3$Mn$_2$Sb$_3$O$_{14}$. La 5$d$ PDOS is multiplied by 20.

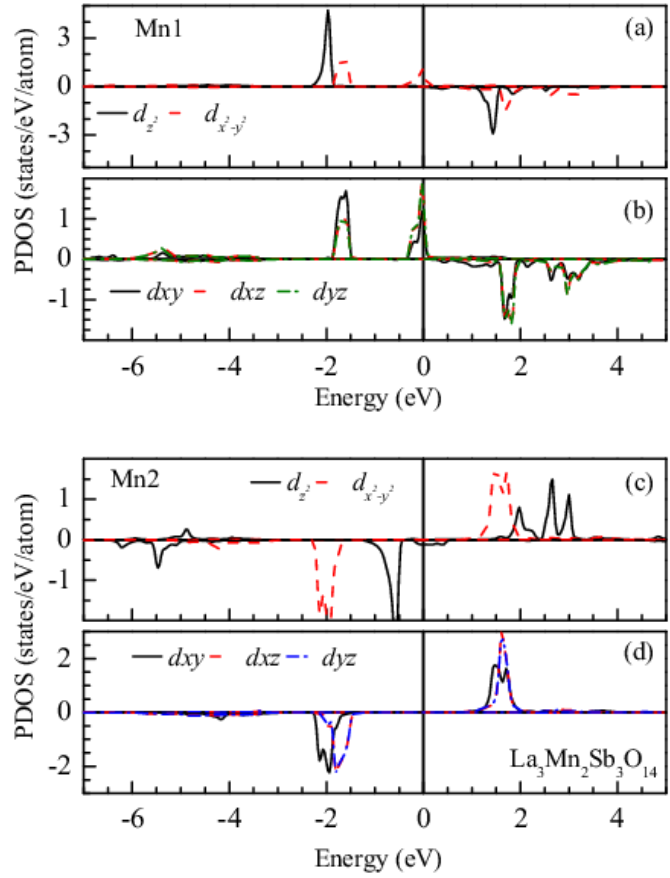

Fig. 7: Partial density of states of Mn 3*d* spin-orbitals. (a) Mn1 $d_{x2-y2}$ and $d_{z2}$, (b) Mn1 $d_{xy}$, $d_{xz}$ and $d_{yz}$, (c) Mn2 $d_{x2-y2}$ and $d_{z2}$, and (d) Mn2 $d_{xy}$, $d_{xz}$ and $d_{yz}$ PDOS.